\documentclass[12pt,preprint]{aastex}

\begin{document}

\title{INFRARED PHOTOMETRY OF THE GLOBULAR CLUSTER PALOMAR~6}

\author{Jae-Woo Lee\altaffilmark{1,2,3}
\& Bruce W. Carney\altaffilmark{1,4}}

\altaffiltext{1}{Department of Physics \& Astronomy,
University of North Carolina, Chapel Hill, NC 27599-3255; 
jaewoo@astro.unc.edu; bruce@physics.unc.edu}
\altaffiltext{2}{Center for Space Astrophysics, 
Yonsei University, Shinchon-dong 134, Seoul 120-749, Korea}
\altaffiltext{3}{PMA Division, California Institute of Technology, 
Mail Stop 405-47, Pasadena, CA 91125, USA; jaewoo@srl.caltech.edu}
\altaffiltext{4}{Visiting Astronomer, Kitt-Peak National Observatory,
National Optical Astronomy Observatory,
which is operated by the Association of
Universities for Research in Astronomy, Inc., under contract with the
National Science Foundation.}

\begin{abstract}
We present $JHK$ photometry of Palomar~6. Our photometric measurements 
range from the RGB-tip to $\approx$ 2 mag below the RHB and our CMDs 
show that Palomar~6 appears to have a well-defined RHB population.
The distance modulus and interstellar reddening of the cluster are estimated
by comparing the magnitude and color of Palomar~6 RHB stars with respect to 
those of 47~Tuc.  We obtain $(m-M)_0$ = 14.28 mag and $E(B-V)$ = 1.30 mag 
for the cluster and our study suggests that Palomar~6 is clearly located 
in the Galaxy's central regions. We also discuss the metallicity of the cluster
using the slope of the RGB. We obtain [Fe/H] $\approx$ $-$1.2 for Palomar~6 and
our metallicity estimate is $\approx$ 0.5 - 1.0 dex lower than 
previous estimates by others.
\end{abstract}


\section{Introduction}
Palomar~6 ($\alpha$ = 17$^h$ 44$^m$, $\delta$ = $-$26$^{\circ}$
13$\arcmin$; $\it{l}$ = 2.1$^{\circ}$, $\it{b}$ = 1.8$^{\circ}$; J2000)
is a globular cluster $\approx$ 0.8 kpc from
the Galactic center and $\approx$ 0.2 kpc from the plane (Harris 1996).
Since it lies near the Galactic center, it is highly reddened.

Malkan (1981) derived the interstellar reddening for Palomar~6, 
$E(B-V)$ = 1.4, based on a reddening-free metallicity index $Q_{IR}$.\footnote{
$Q_{IR}$ is defined to be H$_2$O + 1.5 CO, where H$_2$O and CO are
the narrow-band infrared photometric absorption indices from
H$_2$O at 2.0 $\mu$m~and CO at 2.4 $\mu$m. Malkan (1981) claimed that there
is a correlation between $Q_{IR}$ and a reddening-free ultraviolet
line-blanketing parameter $Q_{39}$ (Zinn 1980) with,
$Q_{IR} = 0.045 + 0.45Q_{39}.$}
Ortolani, Bica, \& Barbuy (1995) obtained $E(B-V)$ = 1.33 $\pm$ 0.10 
based on the $VI$ color magnitude diagram (CMD).
Bica et al.\ (1998) also studied the cluster employing 
intermediate resolution (5.4 \AA~pixel$^{-1}$)
integrated spectra and they derived the smallest value of 
the interstellar reddening value for the cluster, $E(B-V)$ = 1.25. 

The metallicity of the cluster is rather controversial.
Malkan (1981) obtained a metallicity of [Fe/H] = $-$1.30 for Palomar~6
using the $Q_{IR}$ index, ranking it an intermediate 
metallicity inner halo cluster. Zinn (1985) derived [Fe/H] = $-$0.74 
for Palomar~6 by reanalyzing Malkan's photometry.
Ortolani et al.\ (1995) estimated [Fe/H] $\approx$ $-$0.4 based on the
$VI$ CMD. Minniti (1995) studied the metallicities of seven reddened
clusters including Palomar~6. He obtained spectra of six red-giant branch (RGB)
stars in Palomar~6 with a 2 \AA\ resolution covering 
$\lambda\lambda$ 4700 -- 5400 \AA. He suggested that the sum of 
Mg $\lambda$ 5175 \AA, Fe $\lambda$ 5270 \AA, and 
Fe $\lambda$ 5535 \AA\ lines (Mg + 2Fe) is the most sensitive to metallicity
following Faber et al.\ (1985),  and he derived [Fe/H] = +0.2 $\pm$ 0.3 
for Palomar~6 based on the location of the giants 
in the Mg + 2Fe {\em vs} $(J-K)_0$ diagram. More recently, 
Bica et al.\ (1998) obtained [$Z/Z_{\odot}$] = $-$0.09 by measuring
Ca~II triplet $\lambda\lambda$ 8498, 8542, and 8662 \AA\ using 
the intermediate resolution integrated spectra.

As discussed above, Palomar~6 suffers a large interstellar reddening. 
The interstellar extinction in the visual passband
$A_V$ for Palomar~6 is expected to be $\approx$ 3.9 -- 4.3 mag,
depending on its true interstellar reddening value, and 
the apparent visual magnitude of the horizontal branch (HB) 
$V_{HB}$ is $\approx$ 20 mag (Ortolani et al.\ 1995). 
Thus, photometric study of Palomar~6 in the visual passband is very difficult. 
It is also likely that differential reddening, which is commonly 
detected in heavily reddened clusters, is present.
The IR photometry is, therefore, essential for studying Palomar~6, 
since it is less sensitive to the interstellar reddening and 
the differential reddening effect.
In this paper, we explore the ground-based $JHK$ photometry of Palomar~6.
A comparison with $JK$ photometry of Minniti, Olszewski, Rieke (1995)  
is made. The distance modulus using the $K$ magnitude of 
the red horizontal branch (RHB) and the metallicity estimate using 
the RGB slope of the cluster are discussed.

\section{Observations and IR data reduction techniques}
Our observations were carried out using the 4m telescope at KPNO
on the night of June 15 1997 under photometric conditions.
The detector was a 256$\times$256 HgCdTe NICMOS3 array and
the JHK filter system was employed.
The image scale was 0.60 arcsec pixel$^{-1}$ providing a field of view (FOV)
of 2.56 $\times$ 2.56 arcmin.
The journal of observations is presented in Table~1.

During our observations, we dithered the telescope pointing to minimize 
the effects induced by bad pixels and cosmic ray events. 
For standard star frames, the telescope pointing was dithered with 
an offset of 37 arcsec between successive exposures. 
Offsets of 6 and 10 arcsec were applied for Palomar~6 object frames.

NICMOS3 detectors utilize a hybrid architecture
in which each pixel has an associated unit cell 
which controls the biasing and readout of the pixel. 
Thus, each pixel is essentially independent of the others and 
charge bleeding or trailing from saturated pixels is not present. 
However, this independence also means that such
properties as linearity and dark current can vary from pixel to pixel,
and it is necessary to calibrate these effects for optimum scientific
performance.

The mean dark current of NICMOS3 is known to be $\approx$ 2 $e^{-1}$/sec.
However it does not scale simply with the exposure time. 
Thus we obtained dark frames with the same exposure times as object and 
standard star frames to calibrate the dark currents.

The non-linearity of the detector sensitivity was also calibrated.
We obtained a series of dome flats with exposure times of
3, 10, 15, 20, and 30 sec maintaining the same illumination intensity.
We also obtained 1 sec exposures before and after each exposure 
to monitor possible changes in the illumination.
Then the fluxes of each exposure were normalized to that of 1 sec exposures.
The measured fluxes for each exposure were plotted as a function 
of expected fluxes with linear sensitivity.
Then we derived a third-order relation,
\begin{equation}\label{eqn:nicor}
ADU_{corr} = ADU_{obs}[0.9952 + 0.0744 \frac{ADU_{obs}}{32767}
    - 0.0489 (\frac{ADU_{obs}}{32767})^2],
\end{equation}
where $ADU_{corr}$ and $ADU_{obs}$ are the corrected flux and
the observed flux in ADU, respectively. This relation was then applied
to each object frame using the IRAF\footnote{IRAF (Image Reduction and 
Analysis Facility) is distributed by the National Optical Astronomy 
Observatory, which is operated by the Association of Universities for Research 
in Astronomy, Inc., under contract with the National Science Foundation.}
task IRLINCOR. Figure~1 shows the non-linear response curve
of NICMOS3 detector. The solid line represents
the linear relation and filled circles represent the observed fluxes
and open circles represent the corrected fluxes using the relation above.

The sky-flats were generated by median-combining object frames of
the individual dithering sequences for each passband.
This procedure essentially removes the stellar light and cosmic ray,
and provides good sky-flats for less crowded fields, 
such as standard star frames.
The sky-flats generated using Palomar~6 object frames, however, show
residuals of stellar images, indicating that this procedure fails
when the field is too crowded.
Applying sky-flats generated from the other images taken at different times
may cause potential problems.
It is known that there are two predominant sources of infrared sky background,
which are essentially independent, both physically and spectrally.
At short wavelengths (in particular $H$ passband), 
the sky is dominated by emission lines from OH in the upper atmosphere 
(typically 90 km altitude).
The strength of these lines can vary over the course of a night;
in addition, upper level winds create inhomogeneity and motion of the airglow.
As a result, the intensity of the background can vary unpredictably during
the night. At longer wavelengths, thermal emission from the telescope optics
and optically thick telluric lines predominates. The transition between
these two regimes occurs at approximately 2.3 microns, so the background
with filters other than $K^\prime$ or $K$ is primarily OH airglow.
We produced master sky-flats, which were median-combined
sky-flats of the whole set of standard star frames, 
rather than using sky-flats from a single dithering sequence. 
Since the standard star frames were taken over the night in our observation, 
using these master sky-flats will minimize the risk
of the abrupt changes of infrared sky background.

Figure~2 shows a comparison between the sky-flat generated
using the Palomar~6 frame (upper left) and our master sky-flat
(upper right) in the $J$ passband. As can be seen, the sky-flat using the
Palomar~6 frames shows residuals of stellar light.
This causes serious problems in the central part of the cluster,
where the field is more crowded.
The flat-field corrected images are also shown in the Figure.
The lower left panel shows the final image using the sky-flat from
the Palomar~6 frames and the lower right panel shows that of using 
the master sky-flat. The former does not resolve fainter stars very well
and even brighter stars were turned out to be affected by the residual
flux on the sky flat in the central part of the cluster.
Therefore, we used the master sky-flat in our analysis.

It is known that there are several hundred bad pixels in the NICMOS3 detector
used in our observation. The bad pixel correction was applied after 
the flat-field correction step. The bad pixel mask was generated using 
the standard star frames. The flat-field corrected standard star frames 
($\approx$ 100 frames) were normalized and then median combined.
Since the strict definition of a bad pixel is rather subjective
except for the non-responsive (dead) pixels, the pixels deviating more than 
7 $\sigma$ around the mean value (= 1.0) were defined to be bad pixels.
The pixel values at the bad pixel position were interpolated using those of
nearby pixels. Finally, the object frames in each dithering sequence 
were combined. Our combined image of Palomar~6 in $K$ passband
is shown in Figure~3.

During the night, we observed the six UKIRT faint standard stars 
(Casali \& Hawarden 1992).
It should be noted that the number of IR standard stars is 
small and the most of the IR standard stars by Elias et al.\ (1982) are 
too bright for our camera plus telescope configuration.
All standard frames were analyzed using the PHOTOMETRY task in DAOPHOTII
(Stetson 1995). With the results of aperture photometry, 
growth curve analysis was performed using DAOGROW (Stetson 1990) 
to obtain integrated magnitudes.

The extinction coefficients were derived from one standard star (FS25).
The first-order $K$ passband extinction term $k_k$ is
0.101 mag airmass$^{-1}$ and the first-order color extinction terms
$k^{\prime}_{jk}$ and $k^{\prime}_{jh}$ are $-$0.004 and 0.045
mag airmass$^{-1}$, respectively.

To calibrate the photometry, the photometric transformations were assumed
to have the following form,
\begin{eqnarray}
K_{UKIRT} &=& k_0 + \zeta_K, \nonumber \\
(J-K)_{UKIRT} &=& \mu(j-k)_0 + \zeta_{J-K}, \nonumber \\
(J-H)_{UKIRT} &=& \epsilon(j-h)_0 + \zeta_{J-H}.
\label{eqn4:trans}
\end{eqnarray}
The transformation coefficients are $\zeta_K = -3.235 \pm 0.010$,
$\mu = 0.957 \pm 0.019$, $\zeta_{J-K} = 0.624 \pm 0.007$,
$\epsilon = 1.006 \pm 0.016$, and $\zeta_{J-H} = 0.156 \pm 0.004$
in the UKIRT $JHK$ system. (The errors are those of the mean.)
Figure~4 shows the residuals of $K_{UKIRT}$,
$(J-K)_{UKIRT}$, and $(H-K)_{UKIRT}$ using these transformation coefficients.
The mean values  are 0.000 $\pm$ 0.010 in the $K_{UKIRT}$ passband,
0.000 $\pm$ 0.010 in the $(J-K)_{UKIRT}$ color,
and 0.000 $\pm$ 0.007 in the $(H-K)_{UKIRT}$ color.
(The errors are those of the mean.)

Point-spread function (PSF) photometry for all Palomar~6 frames was 
performed using DAOPHOTII and ALLSTAR/ALLFRAME
(Stetson 1987, 1994, 1995; Turner 1995).
The first step was to perform the aperture photometry with star lists
returned from the FIND routine.  We calculated PSFs using isolated 
bright stars. To obtain good PSFs on each frame, at least 3 iterations 
of neighboring-star removal were performed.
For the PSF calculations, we adopted a spatially invariable PSF since, 
unlike modern CCDs, a NICMOS3 array has a small FOV, so that
the stellar radial profiles are not expected to show large variations
over the frame. Also, the number of PSF stars is too small 
(usually 20-30 stars) to calculate reliable variable PSFs in each frame.
In case of 2K CCD system, the number of PSF stars
are large enough (for example 90-110 stars in Lee \& Carney 1999)
to calculate a variable PSF in each frame.
After the first ALLFRAME pass, we undertook an additional pass to find
more stars on the subtracted frames.
Since we used a small PSF radius (usually, $\approx 4\times$FWHM)
to compute the PSF magnitude due to crowding, a PSF magnitude must
be converted into an aperture magnitude with a larger radius,
a process known as {\em aperture correction}. The 20 to 30 PSF stars
on each frame were used for this purpose. Comparisons of the PSF magnitude
and the aperture magnitude from the growth-curve method yielded
aperture corrections. Finally airmass corrections and photometric
transformations were applied.

To convert the UKIRT $JHK$ system to the CIT $JHK$ system, 
we used the following transformation equations given by 
Casali \& Hawarden (1992),
\begin{eqnarray}
K &=& K_{UKIRT} - 0.018(J-K)_{UKIRT}, \nonumber \\
(J-K) &=& 0.936(J-K)_{UKIRT}, \nonumber \\
(H-K) &=& 0.960(H-K)_{UKIRT}.
\label{eqn4:convert}
\end{eqnarray}

\section{Results and discussions}
\subsection{Color Magnitude Diagrams}
Our composite CMDs of Palomar~6 are presented in Figure~5 and 
the sample CMD data are shown in Table 2.
In the Table, the positions (columns 2 and 3) are presented in 
pixel units (0.60 arcsec pixel$^{-1}$).
The CMD data are available upon request to the authors or the
electronic version of the journal.
Off-cluster field populations were not obtained during our observation, 
therefore, the field star contamination is not removed in our CMDs.
Since Palomar~6 lies near the Galactic center, the off-cluster field
contamination is expected to be very high (see below), and 
probably variable due to differential reddening.

Our photometric measurements range from the RGB-tip to $\approx$ 2 mag
below the RHB. The most prominent feature in our CMDs is a well-defined RHB 
morphology at $H \approx$ 13.5 mag and $K \approx$ 13.3 mag. 
The scatter in the RGB sequences appear to be larger in ($J-K$) than
in ($J-H$). This is most likely due to the bright sky background
in the $K$ passband.

\subsection{Comparison with the IR photometry of 
Minniti, Olszewski, \& Rieke} 
Minniti, Olszewski, \& Rieke (1995) obtained $JK$ photometry of Palomar~6
using the 2.3m telescope equipped with a 256$\times$256 HgCdTe NICMOS3 array
at the Steward Observatory.
Their image scale was same as ours (0.60 arcsec pixel$^{-1}$),
but their observations covered wider area (6.6$\times$2.5 arcmin).
To calibrate their photometry, they observed the standard stars of
Elias et al.\ (1982). 
Since standard stars of Elias et al.\ are too bright for
their telescope plus camera configuration, 
they defocused the telescope to prevent the saturation of bright stars
on the detector.
They also observed the globular cluster M22 during the same night and
they made comparisons  with that of Frogel, Persson, \& Cohen (1983);
\begin{eqnarray}
\Delta K_{MOR-FPC} &=& -0.15 \pm 0.11 ~~~~~(n = 3), \nonumber \\
\Delta J_{MOR-FPC} &=& 0.19 \pm 0.15 ~~~~~~~~(n = 4),
\end{eqnarray}
where $\Delta K_{MOR-FPC}$ and $\Delta J_{MOR-FPC}$ refer to
Minniti et al.\ minus Frogel et al.
(The error is 1 $\sigma$ level.)
They noted that the differences are large because their common stars are 
typically bright and close to the nonlinear regime of the detector array.

We compare our photometry with that of Minniti et al.\  
in Figure~6. The differences are in the sense our photometry
minus that of Minniti et al. We compared 134 common bright stars
($K$ $\leq$ 12 mag in our magnitude scale),
and obtained unexpected results.

\noindent
(i) On average, our $K$ magnitudes are 1.384 $\pm$ 0.024 mag brighter
than those of Minniti et al. (The error is that of the mean.) 
There exists a gradient in the difference with a slope of
$\Delta K$ $\propto$ 0.099$\times K$, in the sense that our $K$ magnitudes
are slightly brighter than the mean value for the bright stars.

\noindent
(ii) On average, our $J$ magnitudes are 0.705 $\pm$ 0.021 mag brighter
than those of Minniti et al. There also exists a gradient in
the difference with a slope of  $\Delta J \propto 0.108 \times J$, 
in the sense that our $J$ magnitudes for the bright stars are slightly brighter 
than the mean value.

\noindent
(iii) Our ($J-K$) colors are 0.679 $\pm$ 0.017 mag redder than those of
Minniti et al.

\noindent
The discrepancies between our work and Minniti et al.\ are too large to be 
fully explained by the differences between Minniti et al.\ and Frogel et al.\ 
in M22. We also show CMDs using common stars between our work and 
Minniti et al.\ in Figure~7. In the Figure, the left panel represents 
our photometry and right panel represents that of Minniti et al.\footnote{
Dr. Minniti kindly provided his table for Palomar~6. 
A part of his table is shown below. His $(J-K)$ color values 
in the sixth column are 0.6 mag larger than those from the fifth column 
minus the fourth column. In Figure 7, we adopted the color from 
the fifth column minus the fourth column.
Note that Minniti et al. (1995) used the color in the sixth column 
in their Figure~3. 
\begin{center}
\begin{tabular}{cccccccc}
\hline
X & Y & $r$ & $K$ & $J$ & $(J-K)$ & $\sigma K$ & $\sigma J$ \\
\hline
190.8  & 321.0  & 1.30  & 10.65  & 11.47  & 1.42  & 0.01  & 0.01\\
194.2  & 323.6  & 5.60  & 12.54  & 13.62  & 1.68  & 0.04  & 0.04\\
182.5  & 315.5  & 8.70  & 13.92  & 14.94  & 1.62  & 0.11  & 0.13\\
189.0  & 310.9  & 9.20  & 10.05  & 11.04  & 1.59  & 0.01  & 0.01\\
....... \\
\end{tabular}
\end{center}}

The off-cluster field star contamination is expected to be very high 
toward Palomar~6, as discussed in the previous section, and
it is necessary to know membership RGB stars in our study.
On 23 May 1998, we obtained high resolution 
($\lambda/\Delta\lambda \geq$ 40,000) 
IR echelle spectra of 7 bright RGB stars in Palomar~6
and Arcturus for the $^{12}$CO 2-0 bandhead centered at $\approx$ 22045 \AA\
using the 3.5m NASA Infrared Telescope Facilities.
In order to derive the heliocentric radial velocities, 
we cross-correlated our target spectra with that of Arcturus
($v_r$ = $-$5.2 km sec$^{-1}$, Evans 1967).
Table~3 shows our heliocentric radial velocity measurements for Palomar~6 
RGB stars (see also Figure~3). The third column of the Table shows 
the membership of RGB stars based on their heliocentric radial velocities.
(Hereafter, we refer stars A, C, D, and G as membership RGB stars.)
It should be noted that the previous radial velocity of Palomar~6
is based on the low resolution spectroscopic study of Minniti (1995), 
where the internal measurement error is much larger than ours.
We will discuss the spectroscopic study of Palomar~6 RGB stars
in a forthcoming paper (Lee, Carney, \& Balanchandran 2002, in preparation).

In Figure~8, we show $VJHK$ multi-color CMDs for Palomar~6,
combining our $JHK$ photometry and $V$ photometry of Ortolani et al.\ (1995).
Since we already knew the four cluster membership RGB stars 
(filled circles in the Figure), we drew RGB/RHB fiducial sequences 
with arbitrary widths in the ($V-K$, $V$) CMD by eye 
in the extension of the four membership RGB stars.
We then removed outliers by cross-examining $H-K$ and $J-K$ colors.
The remaining stars (likely the cluster membership stars)
are plotted by crosses in the Figure.
Our ($V-K$, $K$) and ($V-K$, $V$) CMDs may show that
our $JHK$ photometry shown in Figure~5 suffers 
a serious field star contamination,
which is an expected result since Palomar~6 lies near the Galactic center.
The RGB stars are clearly separated into at least 3 distinctive branches
in ($V-K$, $K$) and ($V-K$, $V$) CMDs, in which the cluster's RGB stars
are located to the blue indicating that Palomar~6 is more metal-poor
or less heavily reddened than the surrounding fields.

In Figure~9, we show ($V-I$, $V$) CMDs for 47~Tuc 
(Kaluzny et al.\ 1998) and Palomar~6 (Ortolani et al.\ 1995).
For the 47~Tuc CMD, we show a model isochrone for [Fe/H] = $-$0.83
and [$\alpha$/Fe] = +0.30 (Bergbusch \& VandenBerg 2001) and 
the location of RHB stars in 47~Tuc (see also Figure~34 of VandenBerg 2000). 
For the Palomar~6 CMD, filled circles are for Palomar~6 membership
RGB stars, crosses for stars selected from multi-color CMDs (see Figure~8), 
and gray dots for stars within 1 arcmin from the cluster center. 
Since the model isochrone provides an excellent match with 47~Tuc,
we adopt this model isochrone as a fiducial sequence for 47~Tuc.
We then derived the relative distance modulus and interstellar reddening 
for Palomar~6 with respect to those for 47~Tuc by adjusting 
this model isochrone's magnitude to match with those of 
four Palomar~6 RGB stars (filled circles) and the model isochrone's color 
to match with the color that Ortolani et al.\ (1995) obtained 
at the intersection between the RHB and the RGB.
The crossing point between the two dashed lines centered at $V$ = 19.7 mag 
and $V-I$ = 2.7 mag indicates the RHB magnitude obtained by 
Ortolani et al.\ (1995). The color difference $\Delta (V-I)$ = 1.65 mag 
is corresponding to $\Delta E(B-V)$ = 1.27 mag, assuming $E(V-I) = 1.3E(B-V)$,
and the magnitude difference $\Delta V$ = 5.1 mag corresponds to
$\Delta (m-M)_0$ = 1.16 mag, in the sense of Palomar~6 minus 47~Tuc.
It should be noted that the RHB magnitude reported by Ortolani et al.\ (1995)
appears to be $\approx$ 0.4 - 0.5 mag fainter than that expected from 47~Tuc
(the closed box at $V$ $\approx$ 19.2).
Using these relative distance modulus and interstellar reddening values,
we compare $JK$ photometry for 47~Tuc and Palomar~6 in Figure~10.
The filled circles are for the 47~Tuc RGB/RHB stars (Frogel et al.\ 1981) and
dotted lines at $K$ $\approx$ 6.5 mag represent the RGB-tip $K$ magnitude  
for 47~Tuc (Ferraro et al.\ 2000). The colors and the magnitudes 
of the current work (a) and those of Minniti et al.\ (b)
are shifted by $\Delta (J-K)$ = $-$0.67 mag and $\Delta K$ = 1.60 mag,
assuming $\Delta E(B-V)$ = 1.27 and $\Delta (m-M)_0$ = 1.16 mag between
Palomar~6 and 47~Tuc. In Figure~10-(a), we also show the mean magnitude 
and color of the Palomar~6 RHB stars (see below). 
Our RGB-tip $K$ magnitude appears to be consistent with
that of 47~Tuc, while that of Minniti et al.\ (1995) appears to be
$\approx$ 1.5 mag fainter than 47~Tuc.

A comparison with 2 MICRON ALL-SKY SURVEY (2MASS, Curtri et al.\ 2000) 
may also provide an opportunity to assess our photometry, although 2MASS
may have a potential problem in crowded fields due to a larger pixel
scale (2 arcsec pixel$^{-1}$).
Figure~11 shows a comparison of our photometry with 
that of 2MASS. Since 2MASS employed the $K_S$ passband, 
we converted our magnitudes and colors into those of the 2MASS photometric
system using the transformation relations given by Carpenter (2001).
The results are
\begin{eqnarray}
\Delta K_S &=& -0.010 \pm 0.021, \nonumber \\
\Delta (J-K_S) &=& -0.152 \pm 0.018, \nonumber \\
\Delta (H-K_S) &=& -0.070 \pm 0.020,
\end{eqnarray}
where the difference are in the sense our work minus 2MASS
in the 2MASS photometric system. (The errors are those of the mean.)
The difference in the $(J-K_S)$ color is rather large, in the sense that
our photometry is $\approx$ 0.15 mag bluer than that of 2MASS, but
the differences are much smaller than 1.4 mag in the $K_S$ passband  or
0.7 mag in the $(J-K_S)$ color.

Finally, we observed the four high proper motion stars
(LHS 514, 2584, 2824, and 3094) during the same night.
Table~4 shows a comparison of the unpublished results of
Carney and our measurements.
As can be seen in the Table, our measurements are in good agreement 
with that of Carney to within 0.1 mag.
Therefore we conclude that the discrepancy between our work and
that of Minniti et al.\ (1995) is most likely due to the inaccurate
photometric measurement by Minniti et al.\ (1995).

\subsection{Distance modulus and interstellar reddening}
We derive the distance modulus and the interstellar reddening for Palomar~6
using the method described by Kuchinski et al.\ (1995), 
who recommended to use the magnitude and the color 
at the intersection between the RGB and the RHB in IR photometry.
In Figure~12, we show the $JK$ CMD for 47~Tuc of Frogel et al.\ (1981). 
We also show the slope of the RGB and the mean RHB magnitude and color.
Kuchinski et al.\ (1995) obtained $(J-K)_{(RGB,RHB)}$ = 0.55 mag 
and $K_{(RGB,RHB)}$ = 12.02 mag at the intersection point 
between the RGB and RHB for 47~Tuc.
Following the similar method, we derived these values for Palomar~6,
using the possible membership stars obtained from Figure~8.
First, we derive the slope of the Palomar~6 RGB stars
\begin{equation}
(J-K) =  -0.0751~(\pm 0.0040)\times K +   2.2030~(\pm  0.0452).
\label{pal6slope}
\end{equation}
We then define a rectangle whose sides are parallel to $J-K$ and $K$  
by eye and calculate the mean color and magnitude.
We obtain $\langle J-K\rangle_{RHB}$ = 1.09 $\pm$ 0.04 mag and 
$\langle K\rangle_{RHB}$ = 13.53 $\pm$ 0.15 mag for Palomar~6. 
(The error is 1 $\sigma$ level.) In Figure~12, we show the slope of the RGB, 
the rectangle that we adopted and the mean RHB color and magnitude 
for Palomar~6. We obtain $(J-K)_{(RGB,RHB)}$ = 1.19 mag 
for Palomar~6 using Equation~\ref{pal6slope}.

By comparing the magnitudes $K_{(RGB,RHB)}$ and 
the colors $(J-K)_{(RGB,RHB)}$ at the RGB/RHB intersection points
between 47~Tuc and Palomar~6, we obtained $\Delta E(B-V)$ = 1.21 mag
and $\Delta (m-M)_0$ = 1.09 mag, assuming $E(J-K) = 0.53E(B-V)$ and
$A_K = 0.35E(B-V)$ (Rieke \& Lebofsky 1985).
Using the distance modulus $(m-M)_0$ = 13.25 mag and the interstellar
reddening $E(B-V)$ = 0.04 mag for 47~Tuc (Harris 1996),
we obtained the interstellar reddening $E(B-V)$ = 1.25 mag
and the distance modulus $(m-M)_0$ = 14.34 mag for Palomar~6.
We also derive these values using the $JH$ CMD.
We obtained $(J-H)_{(RGB,RHB)}$ = 0.52 mag and 
$H_{(RGB,RHB)}$ = 12.10 mag for 47~Tuc, and
$(J-H)_{(RGB,RHB)}$ = 0.95 mag and 
$H_{(RGB,RHB)}$ = 13.77 mag for Palomar~6. 
Assuming $E(J-H) = 0.33E(B-V)$ and $A_H = 0.54E(B-V)$ (Rieke \& Lebofsky 1985),
we obtained $E(B-V)$ = 1.34 mag
and  $(m-M)_0$ = 14.22 mag for Palomar~6.
For our study, we adopt the mean values, $E(B-V)$ = 1.30 mag and 
$(m-M)_0$ = 14.28 mag for Palomar~6. 
Our interstellar reddening estimate for Palomar~6 shows good agreement
with previous values. However, our distance modulus estimate is 
$\approx$ 0.5 mag smaller than that of Ortolani et al.\ (1995) who obtained
$(m-M)_0$ = 14.76 using the HB magnitude shown in Figure~8.
Using our distance modulus, the distance of Palomar~6
from the sun is $\approx$ 7.2 kpc.
The Galactocentric distance $R_{GC}$ becomes
0.9 kpc if $R_0$ = 8.0 kpc (Reid 1993).
Palomar~6 is clearly located in the Galaxy's central regions.

Finally, a caution is advised for using RHB magnitudes to estimate 
the distance modulus of the metal-rich globular clusters.
Sandage (1990) studied the vertical height of
the horizontal branch of the globular clusters
as a function of metallicity. His study suggested that
the vertical height of the HB increases with metallicity.
For example, the vertical height of the 47~Tuc RHB is
$\approx$ 0.7 mag in the visual passband. Palomar~6 is expected to
have the similar value of the RHB vertical height (see his Figure~16).
In our CMDs, the Palomar~6 RHB vertical height is $\approx$ 0.5 - 0.6 mag
in the $K$ passband. 
It should also be noted that the $K_0$ magnitude of field red clump stars 
in the Galactic bulge is $\approx$ 13 mag (Alves 2000), 
having similar magnitude to Palomar~6 RHB stars depending on 
the interstellar reddening toward field red clump stars.
Thus accidental inclusion of field red clump stars in our Palomar~6 CMD
could be high and future photometric study of the off-cluster field would 
be desirable.

\subsection{Metallicity}
Kuchinski et al.\ (1995) studied the slope of the RGB in the ($J-K, K$) CMD
as a reddening- and distance-independent metallicity index for metal-rich
globular cluster systems. They derived a linear relation between
metallicity and the slope of the RGB
\begin{equation}
\mathrm{[Fe/H]} = -3.09(\pm 0.90) - 24.85(\pm 8.90)\times(RGB~slope),
\label{rgbslope1}
\end{equation}
where the slope of the RGB is defined to be $(J-K) \propto  K\times(RGB~slope)$. 
Kuchinski \& Frogel (1995) recalculated this linear relation  
with the addition of two more clusters and they obtained,
\begin{equation}
\mathrm{[Fe/H]} = -2.98(\pm 0.70) - 23.84(\pm 6.83)\times(RGB~slope).
\label{rgbslope2}
\end{equation}
We derived the slope of RGB of Palomar~6 and our result is shown in
Figure~11 and Equation~6. As Kuchinski et al.\ (1995) and 
Kuchinski \& Frogel (1995) noted, the slope of the RGB does not depend on 
the interstellar reddening and, therefore, conversion to dereddened colors
and magnitudes is not necessary. With our slope for Palomar~6, we obtained  
[Fe/H] = $-$1.22 $\pm$ 0.18 using Equation~\ref{rgbslope1}
(Kuchinski et al.\ 1995) and $-$1.19 $\pm$ 0.18 using 
Equation~\ref{rgbslope2} (Kuchinski \& Frogel 1995).
It should be noted that our results are from selected Palomar~6 RGB stars
as discussed in the previous section.
If we use all RGB stars, [Fe/H] = $-$1.05 $\pm$ 0.19 
using Equation~\ref{rgbslope1} and $-$1.03 $\pm$ 0.19 using 
Equation~\ref{rgbslope2} and they are slightly lower than those from 
selected RGB stars.\footnote{Referee kindly pointed out that 
the IR HB morphology of Palomar~6 may indicate that its metallicity 
is between M69 ([Fe/H] = $-$0.6) and NGC~6553 ([Fe/H] = $-$0.3) 
from Figure~1 of Ferraro et al. (2000). It should be noted that, however, 
the HB morphology is vulnerable to other effects besides metallicity 
(see for example Lee, Demarque, \& Zinn 1994).}

Our metallicity estimate for Palomar~6 is consistent with that of
Malkan (1981), but does not agree with those of Zinn (1985),
Ortolani et al.\ (1995) and Minniti (1995).
The metallicity of Palomar~6 will be discussed in 
the subsequent paper. 
We obtained [Fe/H] of the three RGB stars in Palomar~6 using 
high-resolution ($\lambda/\Delta\lambda \geq$ 40,000) IR echelle spectra.
Our metallicity of Palomar~6 using IR spectra is 
[Fe/H] = $-$1.08 $\pm$ 0.06, consistent with
those derived from the slope of the RGB.

\section{Summary}
We have presented the $JHK$ photometry of Palomar~6.
Our photometric measurements range from the RGB-tip to
$\approx$ 2 mag below the RHB. 
Our photometry does not agree with that of Minniti et al.\ (1995) and
an independent study using the southern telescope facilities 
would be very desirable in the future.

We have discussed the distance modulus and interstellar reddening
of Palomar~6 by comparing the mean magnitude and color of the RHB stars
with respect to those of 47~Tuc.
Our interstellar reddening estimate is consistent with previous values by
others, while our distance modulus is slightly smaller than that of 
Ortolani et al.\ (1995), who obtained the distance modulus of Palomar~6
by using the similar method in the visual passband.
Nevertheless, our study has suggested that Palomar~6 is clearly located 
in the Galaxy's central regions.

We have also discussed the metallicity of Palomar~6 using the slope of the RGB.
Our metallicity estimate is in good agreement with that of Malkan (1981),
but does not agree with those of Zinn (1985),
Ortolani et al.\ (1995) and Minniti et al.\ (1995).

\acknowledgments
This is part of Ph.D. thesis work of J.\ -W.\ Lee at the University
of North Carolina at Chapel Hill.
J.\ -W.\ Lee thanks Dr.\ Dante Minniti for providing his photometric table.
We also thank an anonymous referee for useful comments and a careful
review of the paper.
This research was supported by the National Aeronautics and Space
Administration (NASA) grant number GO-07318.04-96A from the Space Telescope
Science Institute, which is operated by the Association of Universities
for Research in Astronomy (AURA), Inc., under NASA contract NAS 5-26555 and
the National Science Foundation grants AST96$-$19381 and AST99$-$88156.
Support for this work was also provided in part by the Creative Research
Initiative Program of Korean Ministry of Science and Technology.

\clearpage

\clearpage
\begin{deluxetable}{cccc}
\tablecaption{Journal of observations}
\tablenum{1}
\tablewidth{0pc}
\tablehead{
\colhead{Band} & \colhead{$t_{exp}$} & \colhead{FWHM} & \colhead{Dither} \\
\colhead{} & \colhead{} & \colhead{(arcsec)} & \colhead{(arcsec)}}
\startdata
J & 0.5 $\times$ 5 sec & 1.0 & 6, 10 \\
  & 5.0 $\times$ 5 sec & 1.0 & 6, 10 \\
H & 0.5 $\times$ 5 sec & 1.0 & 6, 10 \\
  & 5.0 $\times$ 5 sec & 1.0 & 6, 10 \\
K & 0.5 $\times$ 5 sec & 1.0 & 6, 10 \\
  & 5.0 $\times$ 5 sec & 1.0 & 6, 10 \\
\enddata
\end{deluxetable}

\clearpage
\begin{deluxetable}{rrrrrr}
\tablecaption{Color-magnitude diagram data}
\tablenum{2}
\tablewidth{0pt}
\tablehead{
\colhead{~~ID} & \colhead{X} & \colhead{Y} & 
\colhead{$K$} & \colhead{$J-K$} & \colhead{$H-K$}}
\startdata
  47 & 186.657 & 16.844 & 10.204 &  1.426 &  0.244 \\
  71 & 237.766 & 25.472 & 10.717 &  1.353 &  0.278 \\
   4 & 224.406 & 28.821 &  8.015 &  1.441 &  0.279 \\
  64 & 196.970 & 46.016 & 10.482 &  1.379 &  0.243 \\
   8 & 258.721 & 49.039 &  8.204 &  1.634 &  0.388 \\
  69 &  29.972 & 52.933 & 10.506 &  1.565 &  0.334 \\
   5 & 249.247 & 62.583 &  7.786 &  1.701 &  0.476 \\
  48 & 165.293 & 69.201 & 10.202 &  1.461 &  0.250 \\
  31 & 174.410 & 73.902 &  9.614 &  1.462 &  0.257 \\
   1 & 221.751 & 80.466 &  7.995 &  1.402 &  0.329 \\
\enddata
\end{deluxetable}

\clearpage
\begin{deluxetable}{ccc}
\tablecaption{Heliocentric radial velocities of Palomar~6 RGB stars.}
\tablenum{3}
\tablewidth{0pc}
\tablehead{
\colhead{ID\tablenotemark{1}} & 
\colhead{$v_r$} & 
\colhead{Membership}\\
\colhead{} &
\colhead{(km/s)} & 
\colhead{} } 
\startdata
A & 185.3 & Yes \\
B & ~26.7 & No \\
C & 173.5 & Yes \\
D & 186.8 & Yes \\
E & $-$13.5 & No \\
F & 134.5 & No \\
G & 176.7 & Yes \\
\hline
Mean & 180.6 $\pm$ 6.5& \\
\enddata
\tablenotetext{1}{See also Figure~3.}
\end{deluxetable}

\clearpage
\begin{deluxetable}{ccc}
\tablecaption{Comparisons of $K$ magnitudes of high proper motion stars}
\tablenum{4}
\tablewidth{0pc}
\tablehead{
\colhead{ID} & \colhead{Carney} & \colhead{This Work}}
\startdata
LHS ~514&~9.19 & ~9.174 $\pm$ 0.011 \\
LHS 2584&12.38 & 12.342 $\pm$ 0.012 \\
LHS 2824&~9.75 & ~9.841 $\pm$ 0.011 \\
LHS 3094&12.25 & 12.216 $\pm$ 0.017 \\
\enddata
\end{deluxetable}


\clearpage
\begin{figure}
\epsscale{1}
\figurenum{1}
\plotone{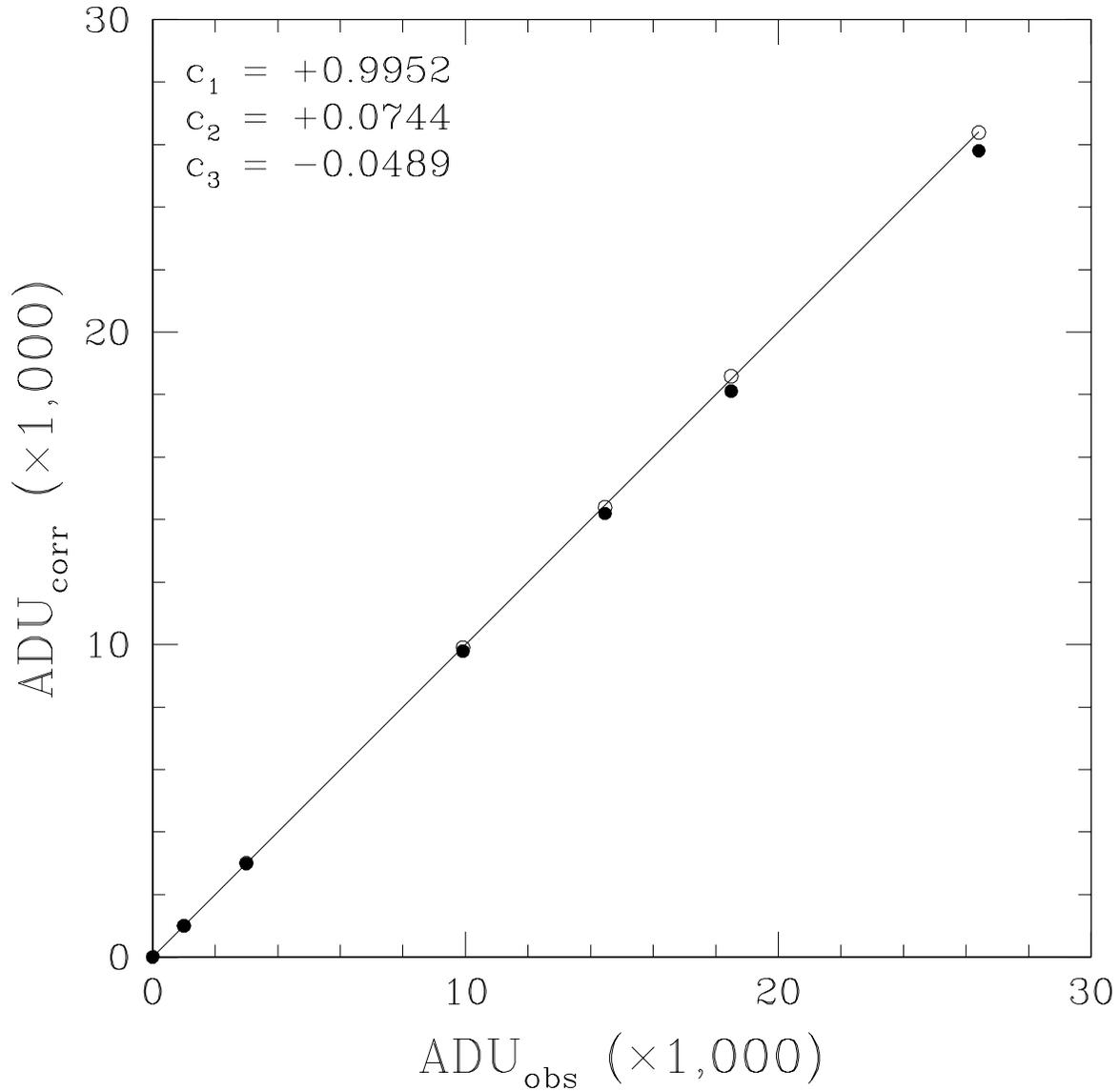}
\caption{The non-linear response of NICMOS3 detector. 
The filled circles represent the observed fluxes and the open circles 
the corrected flux in ADU.
The solid line represents the linear response function.}
\end{figure}

\clearpage
\begin{figure}
\epsscale{1}
\figurenum{2}
\plotone{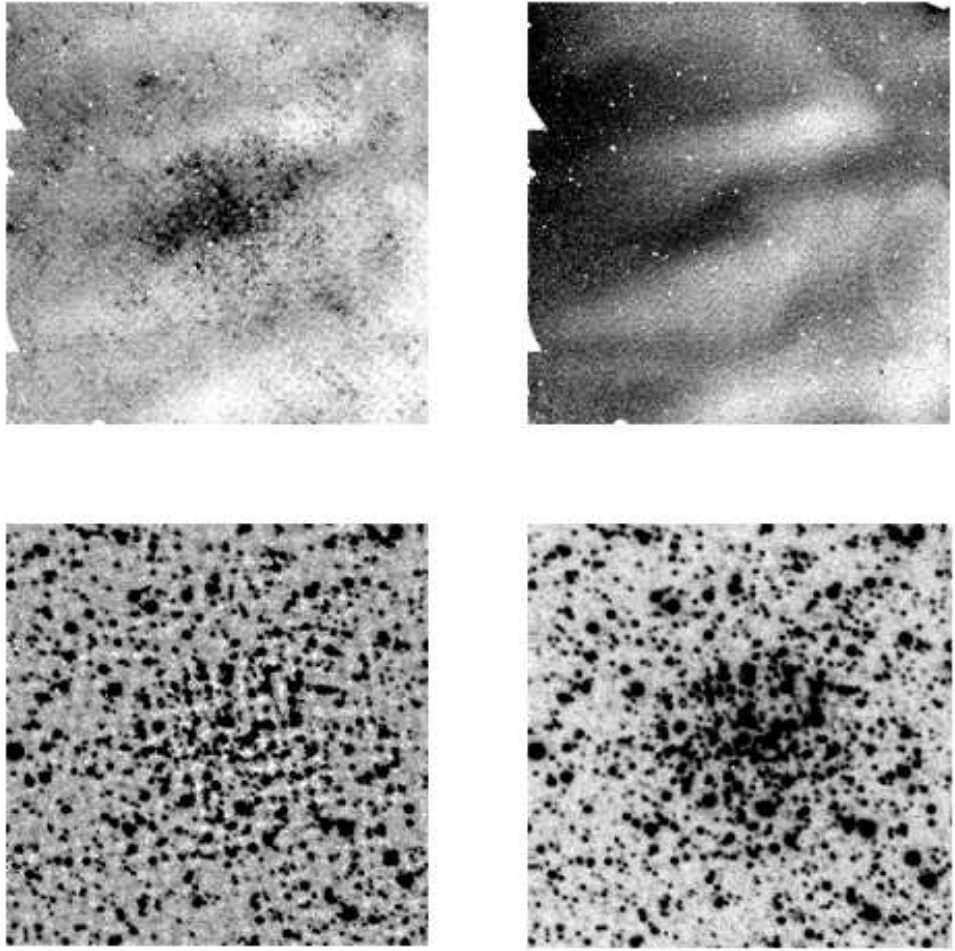}
\caption
{The upper left panel is the sky-flat generated from 
the Palomar~6 object frames and the upper right panel is 
the master sky-flat generated from the standard star frames. 
The upper left panel shows the residual light from
stellar images especially in the central part of the frame.
The lower panels represent the flat-field corrected images for each case.}
\end{figure}

\clearpage
\begin{figure}
\epsscale{1}
\figurenum{3}
\plotone{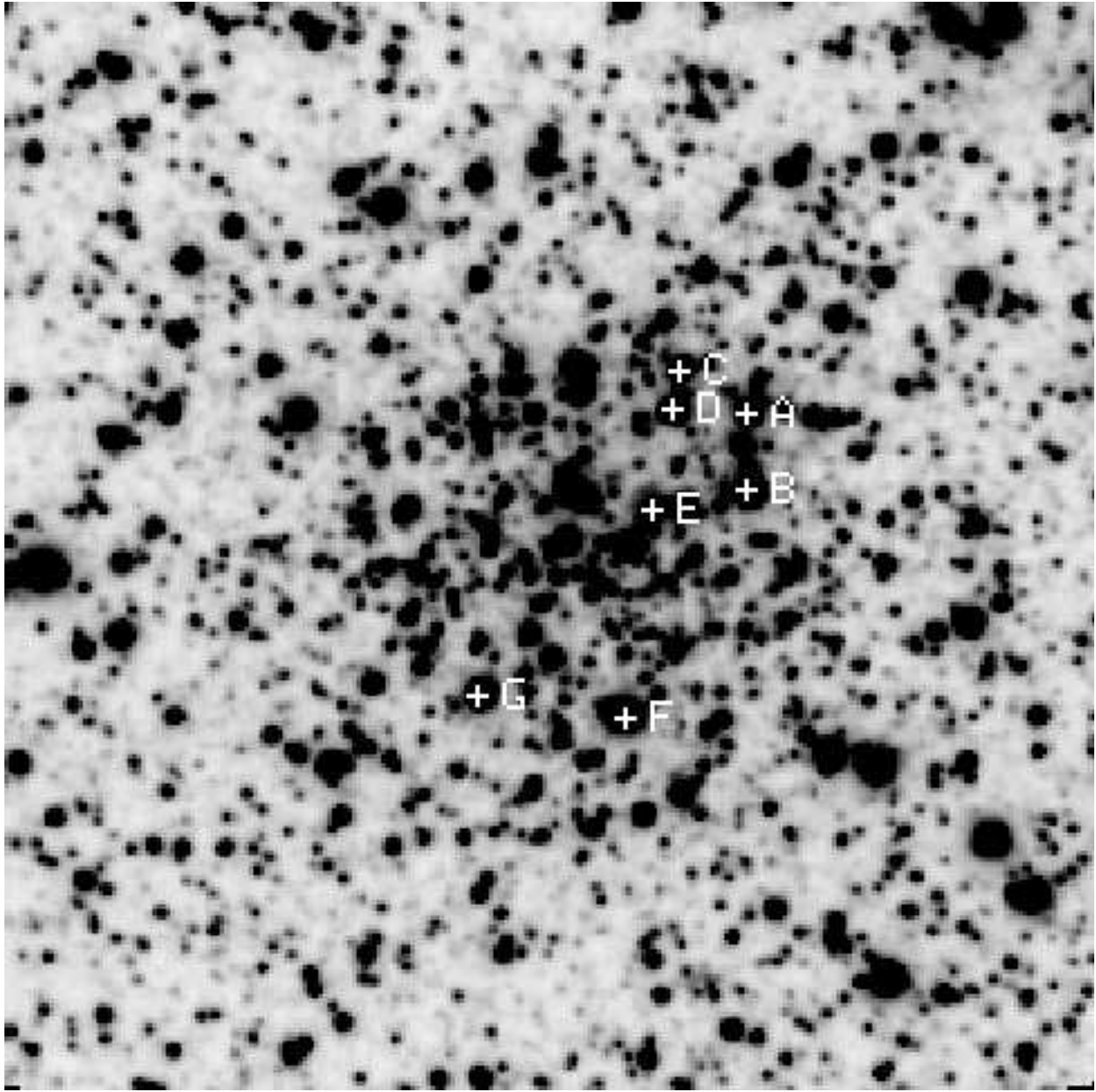}
\caption
{A composite image of Palomar~6 in $K$ passband.
North is at the top and east is to the left.
The RGB stars with known heliocentric radial velocities 
in Table~2 are also marked.}
\end{figure}

\clearpage
\begin{figure}
\epsscale{1}
\figurenum{4}
\plotone{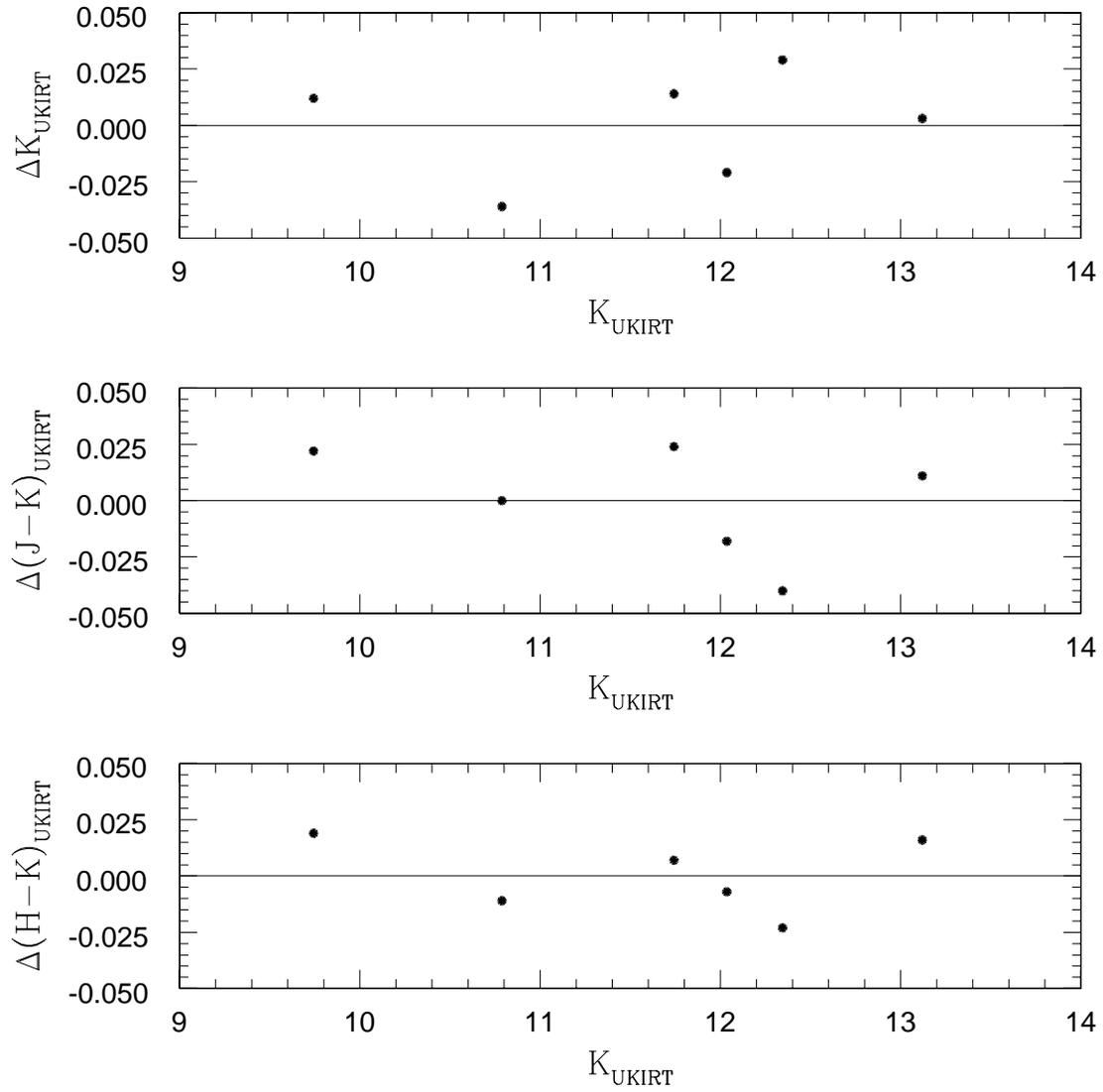}
\caption
{Transformation residuals of the UKIRT faint standard stars.}
\end{figure}

\clearpage
\begin{figure}
\epsscale{1}
\figurenum{5}
\plotone{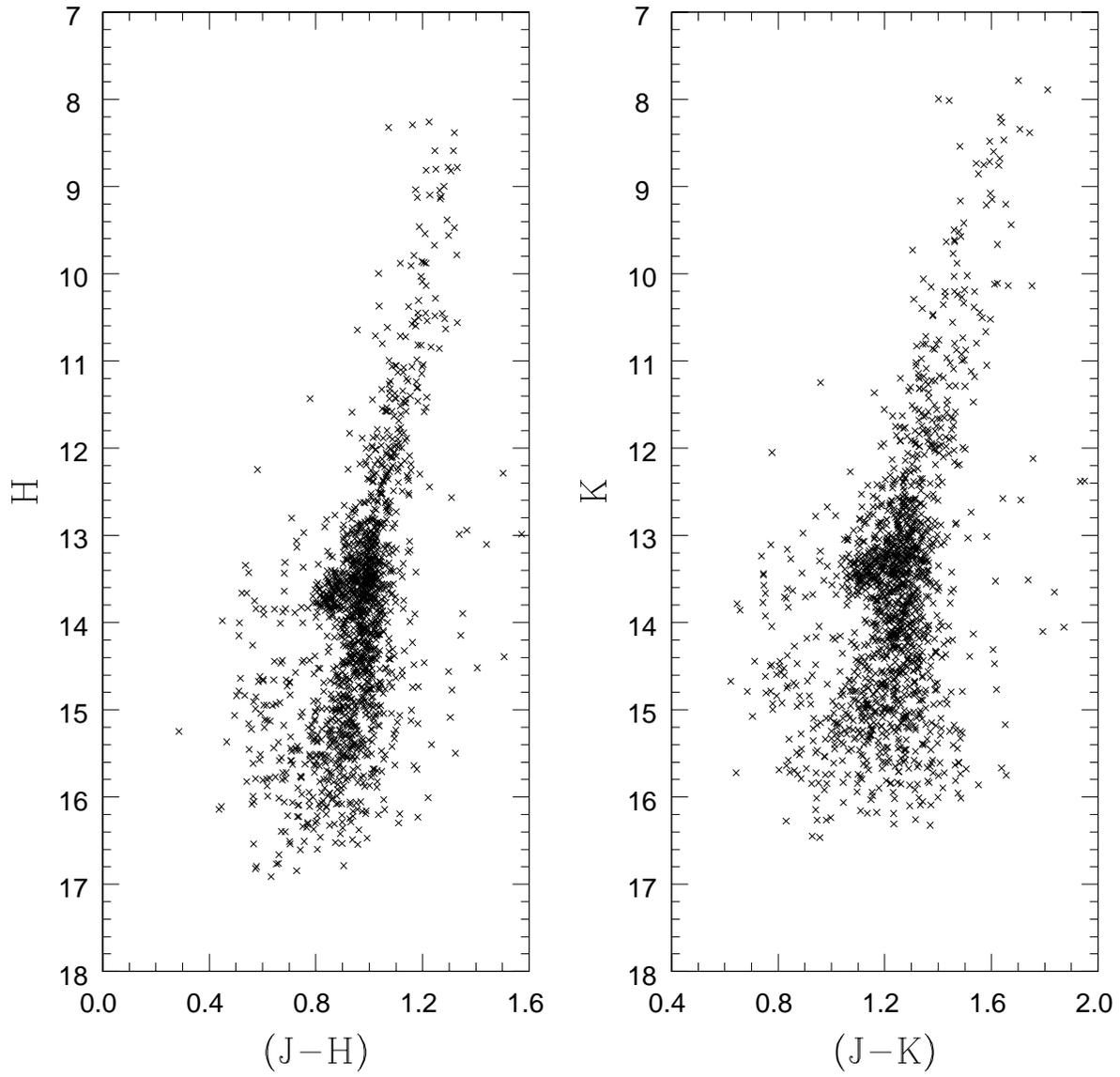}
\caption{Composite IR CMDs of Palomar~6}
\end{figure}

\clearpage
\begin{figure}
\epsscale{1}
\figurenum{6}
\plotone{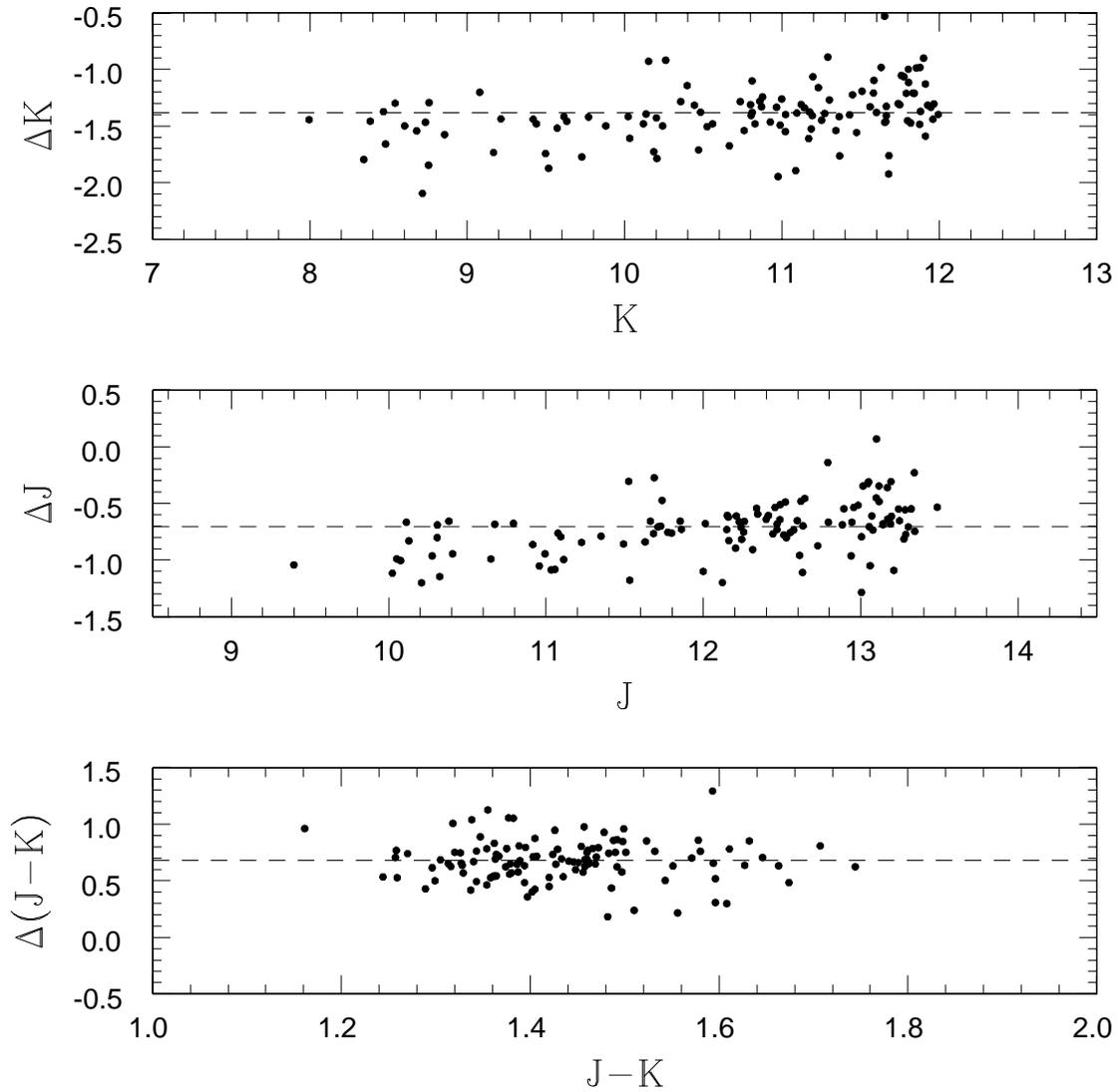}
\caption
{The residuals of the photometry between our work and 
Minniti et al.\ (1995). The differences are in the sense 
our photometry minus that of Minniti et al.}
\end{figure}

\clearpage
\begin{figure}
\epsscale{1}
\figurenum{7}
\plotone{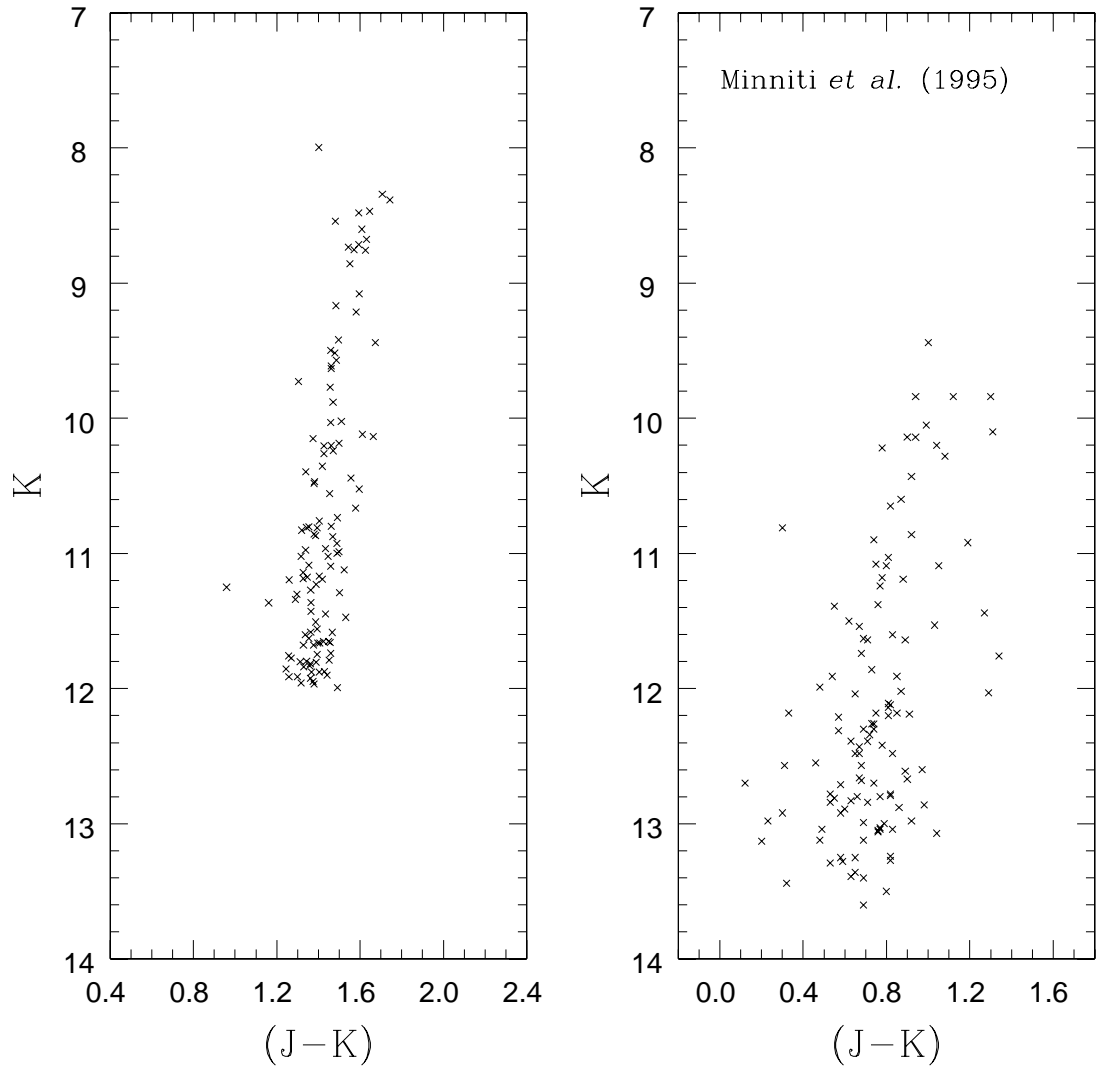}
\caption
{A comparison of CMDs using common stars between our work and Minniti et al.}
\end{figure}

\clearpage
\begin{figure}
\epsscale{1}
\figurenum{8}
\plotone{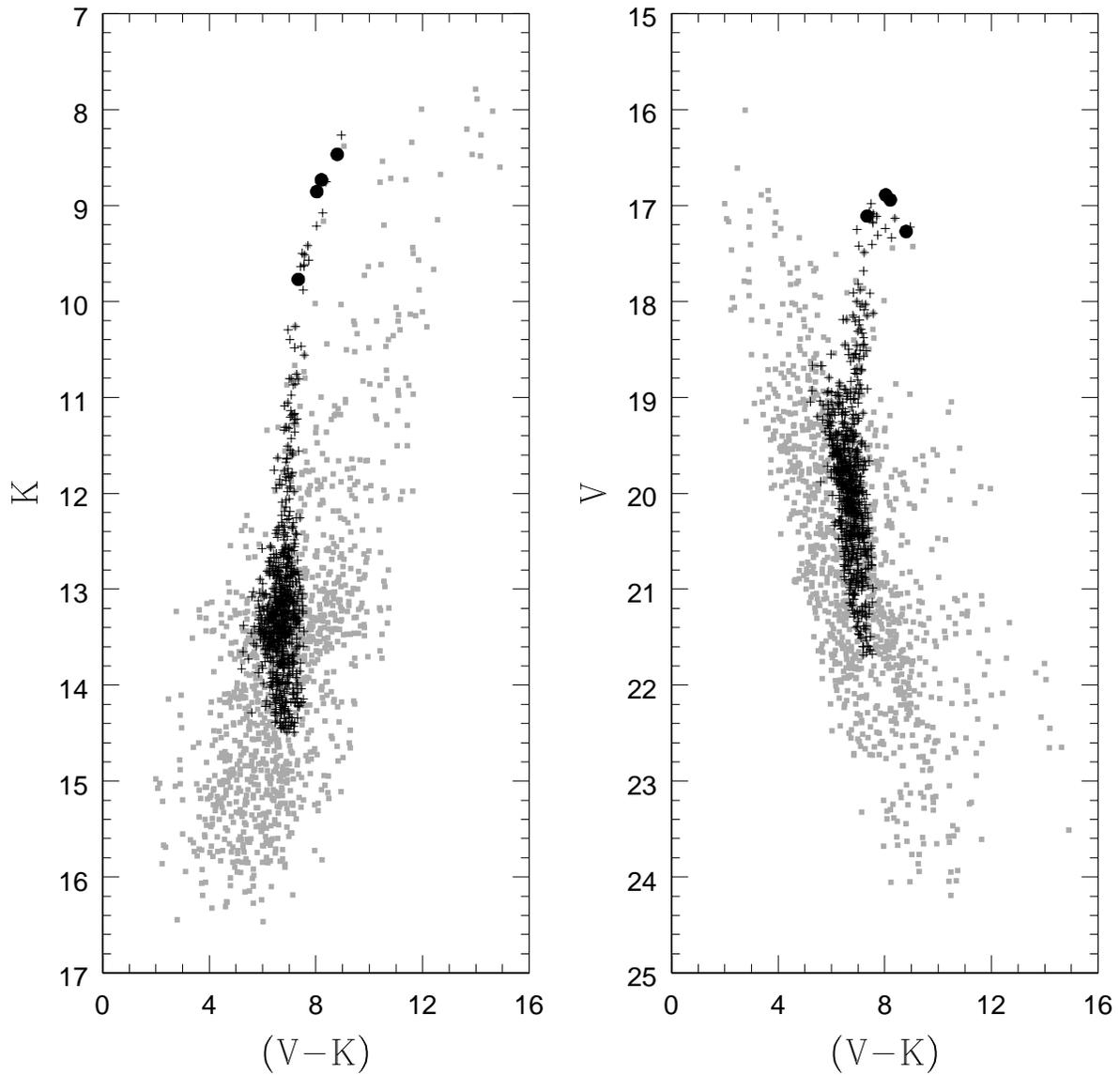}
\caption
{$VJHK$ multi-color CMDs of Palomar~6 and possible membership stars.
Membership stars confirmed by radial velocity measurements are
presented by filled circles and possible membership stars are
presented by crosses.}
\end{figure}
\clearpage
\begin{figure}
\epsscale{1}
\figurenum{8}
\plotone{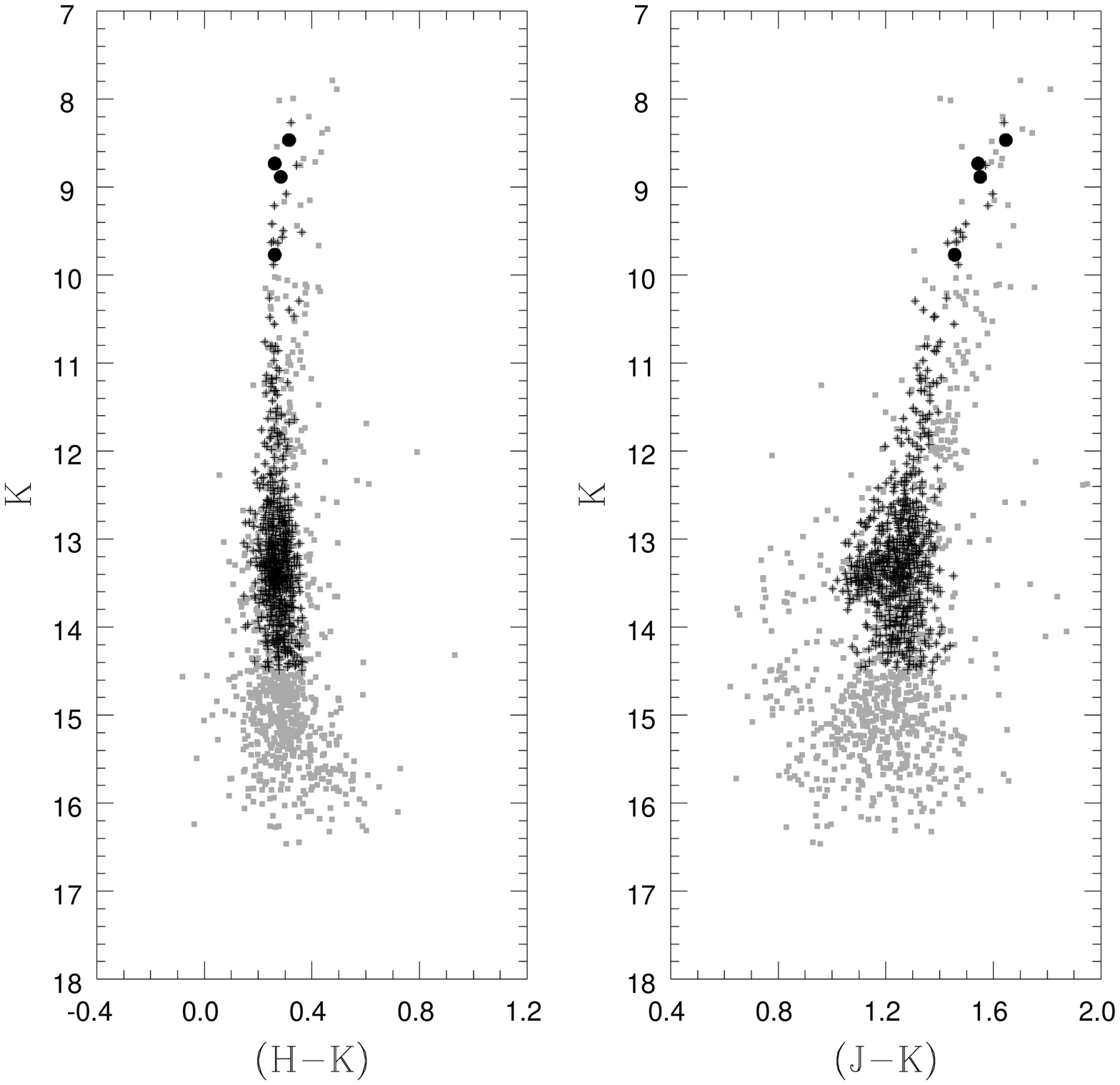}
\caption
{Continued.}
\end{figure}

\clearpage
\begin{figure}
\epsscale{1}
\figurenum{9}
\plotone{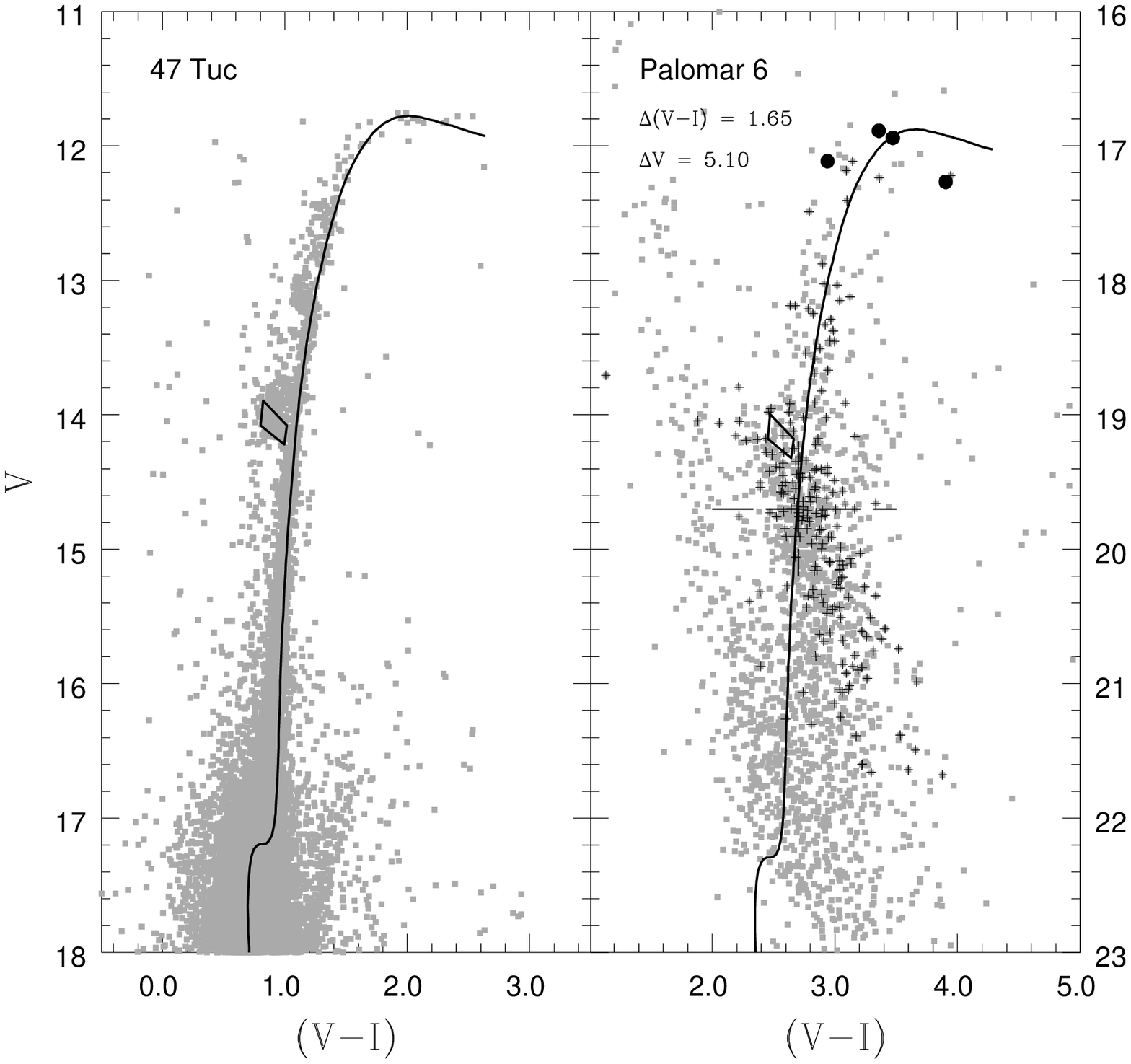}
\caption
{$VI$ CMDs for 47~Tuc (Kaluzny et al.\ 1998) and 
Palomar~6 (Ortolani et al.\ 1995). We show a model isochrone for 
[Fe/H] = $-$0.83 and [$\alpha$/Fe] = +0.30 (Berbusch \& VandenBerg 2001)
and the location of RHB stars in 47~Tuc. 
Using this model isochrone, we derived the relative distance modulus and 
interstellar reddening for Palomar~6 with respect to those for 47~Tuc.
In the right panel, filled circles are for Palomar~6 membership RGB stars,
confirmed by radial velocity measurements, crosses for those selected using
multi-color CMDs (see Figure~8), gray dots for stars within 1 arcmin from
the cluster center, and the crossing point between the two dashed lines 
represnts the RHB magnitude by Ortolani et al.\ (1995).
The color difference $\Delta (V-I)$ = 1.65 is corresponding to
$\Delta E(B-V)$ = 1.27, assuming $E(V-I) = 1.3E(B-V)$, and
the magnitude difference $\Delta V$ = 5.1 mag is corresponding to
$\Delta (m-M)_0$ = 1.19 mag, in the sense of Palomar~6 minus 47~Tuc.}
\end{figure}

\clearpage
\begin{figure}
\epsscale{1}
\figurenum{10}
\plotone{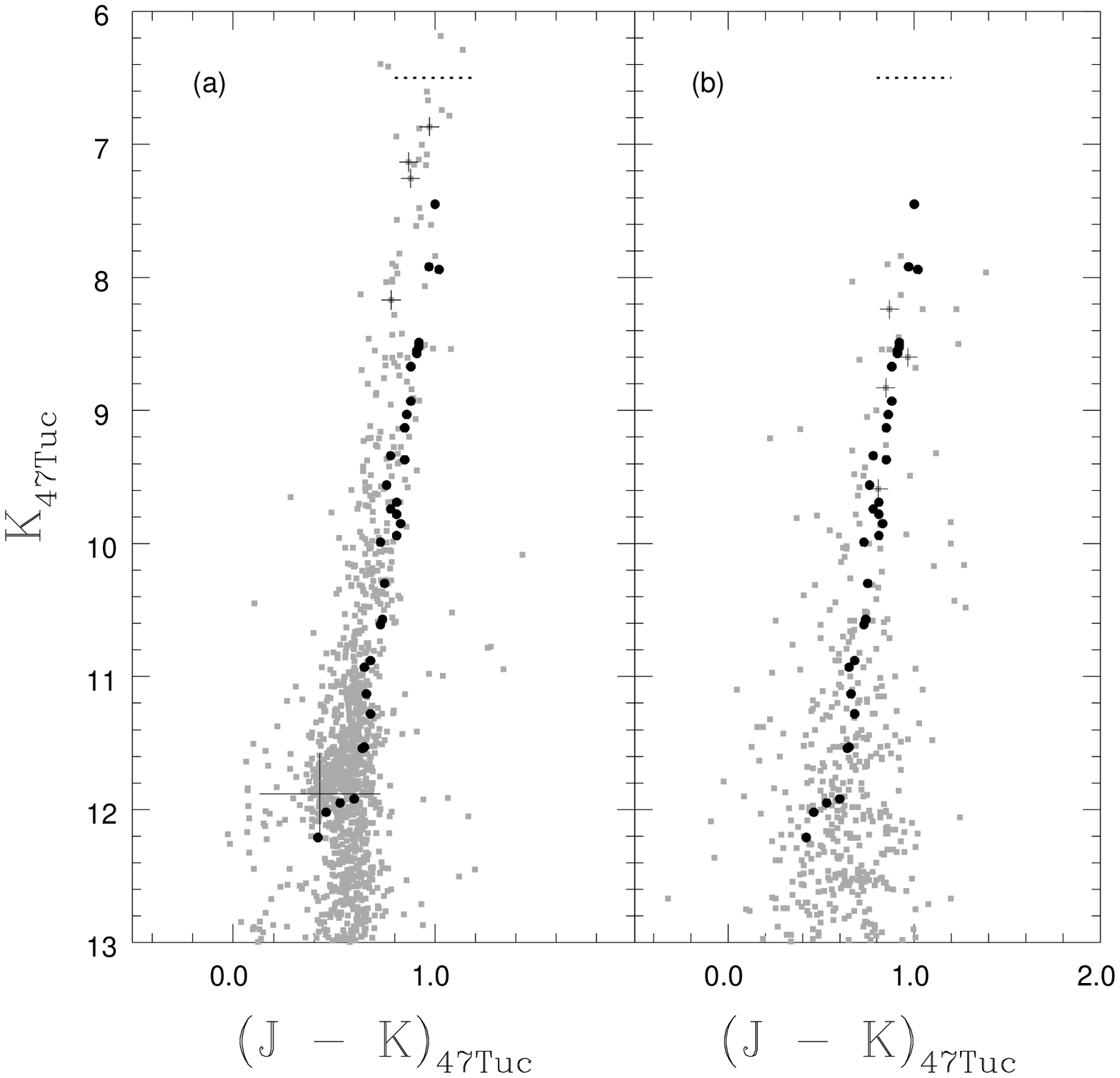}
\caption
{A comparison of $JK$ photometry of Palomar~6 with respect to 
that of 47~Tuc. The filled circles are for the 47~Tuc RGB/RHB stars 
(Frogel et al.\ 1981) and dotted lines at $K$ $\approx$ 6.5 mag show 
the $K$ magnitude of the brightest RGB stars in 47~Tuc 
(Ferraro et al.\ 2000). The colors and the magnitudes
of the current work (a) and those of Minniti et al.\ 
(b) are shifted by $\Delta (J-K)$ = $-$0.67 mag and $\Delta K$ = 1.60 mag,
assuming $\Delta E(B-V)$ = 1.27 and $\Delta (m-M)_0$ = 1.16 mag between
Palomar~6 and 47~Tuc. In Figure (a), we show the mean magnitude and color
of the Palomar~6 RHB stars. The four membership RGB stars are represented
by crosses in each panel. Our RGB-tip $K$ magnitude is consistent with
that of 47~Tuc, while that of Minniti et al.\ is $\approx$ 1.5 mag
fainter than 47~Tuc.}
\end{figure}

\clearpage
\begin{figure}
\epsscale{1}
\figurenum{11}
\plotone{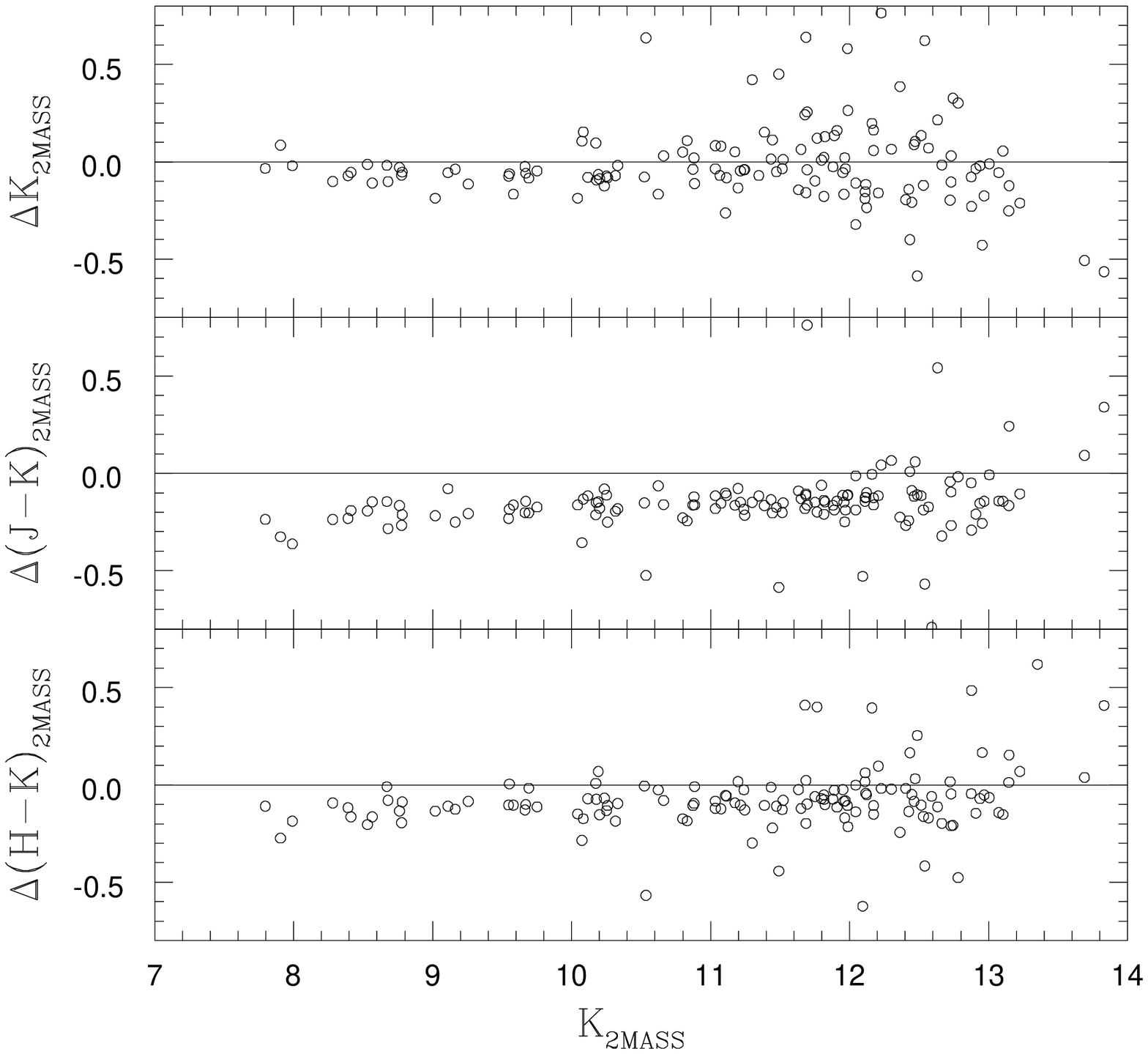}
\caption
{A comparison with 2MASS photometry. The differences are 
in the sense our photometry minus that of 2MASS.}
\end{figure}

\clearpage
\begin{figure}
\epsscale{1}
\figurenum{12}
\plotone{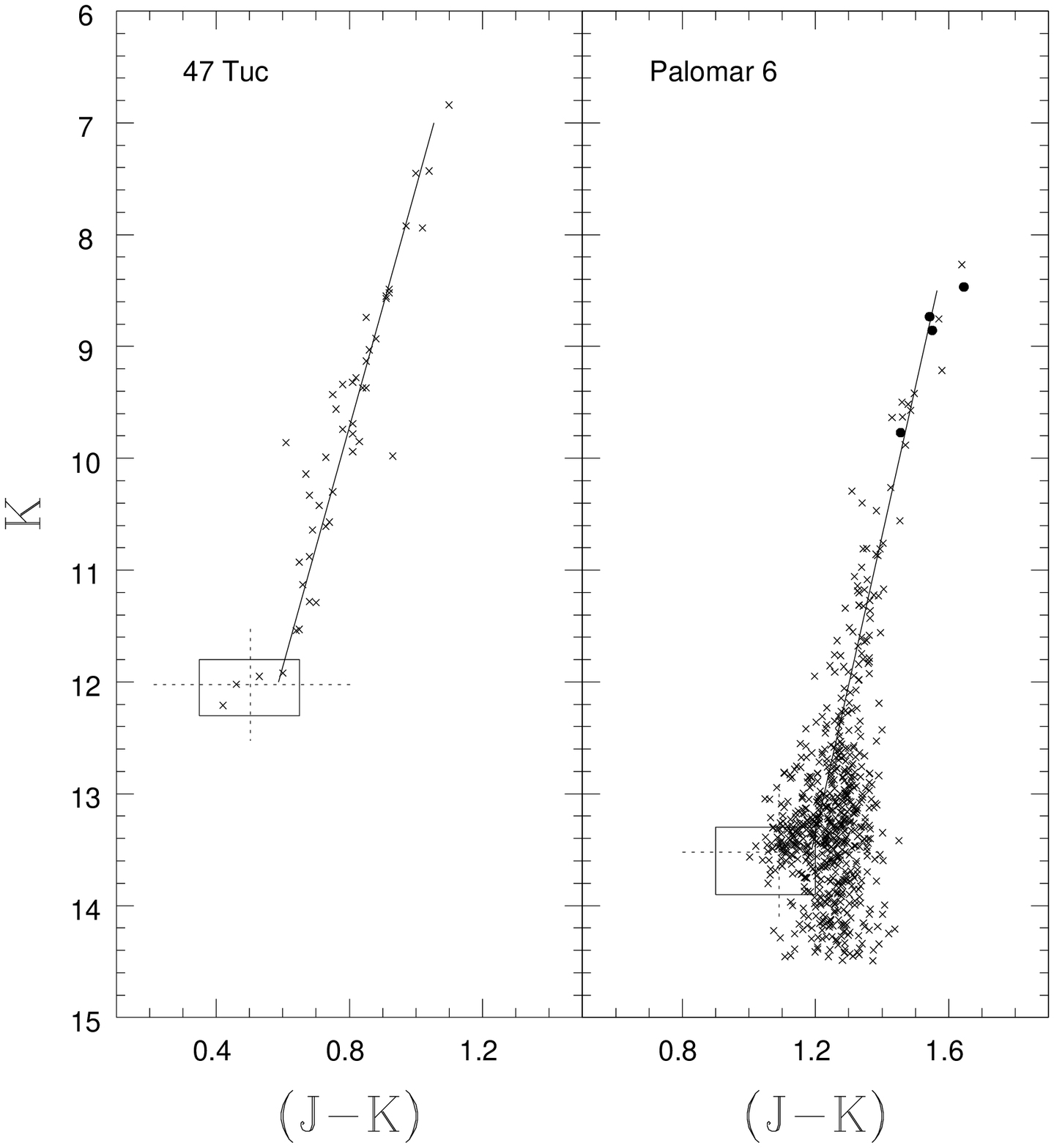}
\caption
{The mean RHB colors and magnitudes for Palomar~6 RHB and 47~Tuc.
The center of the dotted lines represents the mean RHB color and magnitude,
using stars inside the rectangle. The filled circles represent the Palomar~6
RGB membership stars.}
\end{figure}

\end{document}